\documentclass[12pt,epsf]{article}

\usepackage{graphicx}
\usepackage{mathrsfs}
\usepackage{amssymb}
\usepackage{amsmath}

\newcommand{\be}{\begin{equation}}

\newcommand{\ee}{\end{equation}}

\newcommand{\ba}{\begin{array}}

\newcommand{\ea}{\end{array}}

\begin{document}
\begin{titlepage}
\vspace{.5in}
\begin{flushright}
CQUeST-2013-0634
\end{flushright}
\vspace{0.5cm}

\begin{center}
{\Large\bf The nucleation of false vacuum bubbles with compact geometries}\\
\vspace{.4in}

  {$\rm{Bum-Hoon \,\, Lee}^{\S\dag}$}\footnote{\it email:bhl@sogang.ac.kr}\,\,
  {$\rm{Chul \,\, H. \,\, Lee}^{\ddag}$}\footnote{\it
  email:chulhoon@hanyang.ac.kr}\,\,
  {$\rm{Wonwoo \,\, Lee}^{\S}$}\footnote{\it email:warrior@sogang.ac.kr}\, \,
  {$\rm{Changheon \,\, Oh}^{\P}$}\footnote{\it
  email:choh0423@gmail.com} \\

  {\small \S \it Center for Quantum Spacetime, Sogang University, Seoul 121-742, Korea}\\
  {\small \dag \it Department of Physics, Sogang University, Seoul 121-742, Korea}\\
  {\small \ddag \it Department of Physics, Hanyang University, Seoul 133-791, Korea}\\
  {\small \P \it Nost Strategy Institute, Gyeonggi-do 431-904, Korea}\\

\vspace{.5in}
\today
\end{center}

\begin{center}
{\large\bf Abstract}
\end{center}
\begin{center}
\begin{minipage}{4.75in}

{\small \,\,\,\,We investigate and classify the possible types of false vacuum bubbles in Einstein gravity. The false vacuum bubbles can occur only if gravity is taken into account. We show that there exist solutions only with compact geometries. The analytic computations for the radius and nucleation rate of a vacuum bubble are evaluated using the thin-wall approximation. We discuss possible cosmological implications of our new solutions.}


\end{minipage}
\end{center}
\end{titlepage}



\newpage
\section{Introduction \label{sec1}}

The inflationary multiverse scenario was suggested as a nonsingular regenerating inflationary universe, which contains an infinite number of mini-universes or bubbles with all possible vacuum states realized \cite{alinde00, alinde01}. Our universe would correspond to one of the mini-universes or bubbles. According to that scenario, there might seem to be no beginning of the whole universe. The expanding bubble as our universe was introduced in the $1930$s \cite{edding}, in which the bubbles were defined as causally disconnected regions between which no causal influence can ever pass. In this paper, we are interested in finding the new types of false vacuum bubbles in Einstein gravity, which will turn out to have compact geometries.

The bubble nucleation may give some clue to the following three questions:
\begin{itemize}
\item Whether the universe could avoid a initial singularity or not?
\item Whether the universe came from something or nothing? and
\item How could the universe be created with a very low entropy state?
\end{itemize}

The standard cosmology has the initial big-bang singularity problem \cite{hwkpen}. This implies the incompleteness of causal geodesics in the past direction. The singularity theorem implies that a spacetime can not satisfy causal geodesic completeness if, together with Einstein's equations in the presence of a matter source, some conditions hold \cite{hwkpen}. One of traditional prescriptions for avoiding this problem is to hope the quantum gravity to resolve the initial singularity problem. The other prescription for this problem is to violate one condition among the assumptions of the theorem. However, it does not guarantee that the initial singularity could be avoided.

Can the inflationary cosmology avoid this problem? During the inflation the universe violates a suitable energy condition, one of the key ingredients in various singularity theorems. Moreover, many inflating spacetimes are likely to violate the weak energy condition by quantum fluctuations \cite{bv06}. Unfortunately, it turns out that the inflationary cosmology suffers from the initial singularity problem \cite{bvg}. The criteria of their analysis is that the average rate of the Hubble expansion in the past is greater than zero. The simplest strategy is to find the scenario having the rate less or equal to zero in the past. One of these scenarios is known as the emergent universe. The universe was asymptotically the closed Einstein static universe at the past infinite, i.e.\ the universe has no initial singularity \cite{emt, emt01}. However, the scenario has instability \cite{guen}, which means that the universe in this scenario has the finite history in the past direction.

We make short comments on the de Sitter (dS) spacetime. The dS spacetime as the spatially flat Friedmann model is geodesically incomplete. While dS spacetime as the closed Friedmann model can be geodesically complete, which means the coordinates cover the whole space \cite{hael}. Inflation begins after reaching its minimum radius. Thus the necessary part for inflation is only half of the dS. In this regard, dS for inflation is incomplete in the past direction. A different point of view avoiding the singularity problem using two arrows of time separated by a dS like bounce was studied in Ref.\ \cite{agga} and \cite{cache}, in which the special boundary conditions at the bounce are required \cite{avil78}.

There have been many studies on this subject including various theories of gravity \cite{mboj, mboj01, gava, gava01, gava02, gava03}, though not completely known. One could guess that a nonsingular model in those theories may correspond to the model with the violation of the energy condition in the view point of Einstein gravity. Recently the dynamical transition from anti-de Sitter(AdS) to dS avoiding big crunch singularities was studied in Refs.\ \cite{gusi, gusi01}. However, this transition is not connected to the nucleation process of a vacuum bubble.

If we consider the multiverse rather than a single universe, the initial singularity issue seems to be not clear. In the chaotic inflationary model, different parts of the expanding region was created at different moments of time, and then grow up. Thus the universe as a whole does not have a single beginning \cite{linde99}. This scenario may have the so-called measure problem, which we will not mention further in the present paper.

We now address the case of creating a universe. In Ref.\ \cite{tryon}, the author speculatively considered our universe appeared as a fluctuation of the vacuum. He considered our universe is closed one. In order to obey the energy conservation, he employed the well-known fact that the total energy of a closed universe is vanishing. In his model, he did not explain about the origin of the background universe.

There have been studied on the creation of a universe in the laboratory \cite{bgg06, bgg07, bgg08}. The authors considered the massive false vacuum bubble over some critical value of the mass. To the outside observer it is surrounded by a black hole. For the inside observer, the false vacuum bubble can inflate without eating up the true vacuum region. The outside observer can only see the black hole and can not recognize the creation of a universe. In Refs.\ \cite{sato80, sato81} this disconnecting region was called as a child universe. However, these bubbles start from the initial singularity or have special boundary conditions (see \cite{rubakov} for the recent work). We tried to obtain the false vacuum bubble within the infinite geometry as the nucleation process of a vacuum bubble in Ref.\ \cite{llqw00}, which has error terms, thus the question within the reasonable parameter ranges remains to be explored. The case corresponding to the small dS false vacuum bubble with the negative tension within the large dS geometry (compact geometry) is possible (The case is not possible in the Einstein gravity.) \cite{llqw00, kllly}.

A simple question on the creation of the universe is whether our universe was created from something or nothing, in which something means that there are a spacetime as wall as the laws of physics forever. And there was a event correspond to the creation of a universe without any singularity through certain uncertain mechanism from the parent environment. Nothing means that the universe created with a minimum size from nothing as a quantum event \cite{vil07, vil08, vil09, vil10} or through the no boundary proposal \cite{haha}. In this respect, nothing would be unstable. One could assume the laws of physics could be applied into the nothing to treat the creation as a physical event.

The cosmic landscape scenario is the design that involves a huge number of different metastable and stable vacua providing all possible states realizing the laws and constants of nature \cite{boupo, kkk01, kkk02, kkk03}. In this kind of framework the tunneling process becomes a remarkable event. If the vacuum has a positive energy density, dS, then the transition from a lower to a higher vacuum is possible. This is the so-called recycling process \cite{gavil00, gavil01}. On the other hand, if the vacuum has a negative or zero energy density, the reverse tunneling from the true to the false vacuum state is not possible. Those are called as terminal vacua \cite{gsvw}, sinks \cite{linde00}, or black region \cite{sss00}. The fraction of the comoving volume of dS vacua decreases as time goes on due to the existence of the sinks \cite{linde00}. The black region has vanishing cosmological constant as a non-inflating region. They described that the rate for tunneling from the black vacuum to white vacuum with the positive vacuum state is vanishing \cite{sss00}. If the string theory landscape scenario allows our solutions, then the concept of above terminologies could be modified due to the nucleation of false vacuum bubbles.

The early universe or inflationary phase may begin with the extraordinarily low entropy state \cite{penrose1}. One does not know the exact reason why the universe began with the very low entropy state. In this respect, the low entropy state in the early universe could be an additional assumption as an initial condition. If our universe would be a kind of fluctuation out of an equilibrium state in the eternal spacetime, the ordinary observer like us would be much less than the total number of Boltzmann brains. This is the Boltzmann brain problem \cite{also9, also91, also92}. To avoid the problem, we should show that the universe is not a random fluctuation out of an equilibrium state or the entropy of the universe is unbounded from above. In the inflationary universe scenario, the entropy could be produced in the reheating process, the transition period from the dS like phase to Friedmann universe, in which the process corresponds to a highly complicated non-equilibrium period. When one ignores the production or the existence of black holes \cite{Khlopov, Khl10}, almost of the total entropy of the present observable universe were created during the reheating phase according to the inflationary scenario \cite{guthin, frhs}.  The entropy puzzle for the universe with randomly-chosen initial condition \cite{cache, cache0, cache01} could be naturally avoided by the above fact, i.e.\ beginning with the very low entropy state and the highly complicated irreversable events.

According to the recent measurements, one of largest contributors to the entropy of the observable universe today is supermassive black holes at the centers of galaxies $S_{SMBH} \varpropto 10^{104}$. The entropy within the cosmic event horizon as the largest contributor is estimated as $S_{CEH} \varpropto 10^{122}$ \cite{egli}. Recently the entropy of the universe was introduced as an entanglement entropy \cite{dongsu}. How could the entropy of the universe be unbounded from above?

Based on the above arguments, the tunneling process giving rise to the nucleation of a vacuum bubble may take on renewed importance. The purpose of this paper is to investigate the new types of false vacuum bubbles in Einstein gravity and their cosmological implications. The false vacuum bubbles with only compact geometries will be shown to be allowed. We will classify vacuum bubbles according to the size of dS regions and the sign of the cosmological constant of the each vacuum region. The nucleation of a vacuum bubble without gravity was studied in Refs.\ \cite{kov, col002} and was extended in curved spacetime \cite{bnu02, par02}. A homogeneous Euclidean configuration in which the scalar field jumps simultaneously onto the top of the potential barrier was studied in Ref.\ \cite{hamo}. The mechanism for the nucleation of a false vacuum bubble within the true vacuum background has also been studied. Nucleation of a large false vacuum bubble in dS space was originally obtained in Ref.\ \cite{kw04}. In the present work, we extend possible types of false vacuum bubbles to flat and AdS background in Einstein gravity. The nucleation of a false vacuum bubble with flat or AdS geometry could provide the scenario with no initial singularity and coming from something. For the entropy issue, the scenario by the nucleation of a false vacuum bubble could also provide the beginning with a zero or very low entropy state. If the nucleation process could occur occasionally then the the entropy of the given background could be unbounded from above by the irreversable events. In the semiclassical approximation, the nucleation probability of the vacuum bubble is given by $\Gamma \varpropto Ae^{-B/\hbar}$ \cite{col002}, in which the pre-exponential factor $A$ was studied in Refs.\ \cite{cacol, cacol01, cacol02, cacol03, cacol04}. The exponent $B$ is the difference between Euclidean action corresponding to the solution and that of the background geometry. We are interested in obtaining the coefficient $B$.

The paper is organized as follows: In the next section we set up the boundary conditions for our new solutions with compact geometries. We employ the Euclidean approach to obtain the action providing the probability for tunneling. We classify the possible types of vacuum bubbles (true vacuum bubbles, the bounce solutions for degenerate vacua, and false vacuum bubbles). It is argued that false vacuum bubbles have only compact geometries in Einstein gravity. In Sec.\ \ref{sec3}, we numerically solve the coupled equations for the gravity and the scalar field. We show that there exist the new solutions representing the false vacuum bubbles with the compact geometry. We evaluate the radius of bubbles and the probability of the solutions using the thin-wall approximation. In the final section, we summarize our results and discuss the cosmological implications.

\section{Set-up and the classification of vacuum bubbles \label{sec2}}

We study the simple model in the Einstein gravity with a minimally coupled scalar field. The scalar field potential has two non-degenerate minima. One minimum corresponds to a true vacuum state and the other corresponds to a false vacuum state. Both states could be spatially homogeneous and classically stable. However, they could be metastable quantum mechanically, and so they could decay {\it via} tunneling processes. We study an inhomogeneous tunneling channel.

Let us consider the following action:
\begin{equation}
S= \int_{\mathcal M} \sqrt{-g} d^4 x \left[ \frac{R}{2\kappa}
-\frac{1}{2}{\nabla^\alpha}\Phi {\nabla_\alpha}\Phi
-U(\Phi)\right] + \oint_{\partial \mathcal M} \sqrt{h} d^3 x
\frac{K-K_o}{\kappa}\,, \label{f-action}
\end{equation}
where $\kappa \equiv 8\pi G$, $g\equiv det g_{\mu\nu}$, $R$ denotes  the Ricci curvature of spacetime $\mathcal M$, $K$ and $K_o$ are traces of the extrinsic curvatures of ${\partial \mathcal M}$ computed with outward-pointing normals in the metric $g_{\mu\nu}$ and $\eta_{\mu\nu}$, respectively, and the second term in the right-hand side is the so-called York-Gibbons-Hawking (YGH) boundary term \cite{ygh, ygh01}. The gravitational field equations can be obtained properly from a variational principle with this boundary term. This term is also necessary to obtain the correct action. Here we adopt the notations and sign conventions in Ref.\ \cite{mtw}.

The scalar potential $U(\Phi)$ in Eq.\ (\ref{f-action}) has two
non-degenerate minima with the lower minimum at $\Phi_T$ and the higher minimum at $\Phi_F$
\begin{equation}
U(\Phi) = \frac{\lambda}{8} \left(\Phi^2-
\frac{\mu^2}{\lambda}\right)^2 -\frac{\epsilon \sqrt\lambda}{2\mu}\left(\Phi-\frac{\mu}{\sqrt{\lambda}} \right) + U_o\,, \label{poten}
\end{equation}
where the parameter $\epsilon$ is roughly the difference between $U(\Phi_F)$ and $U(\Phi_T)$.

\begin{figure}
\begin{center}
\includegraphics[width=4.5 in]{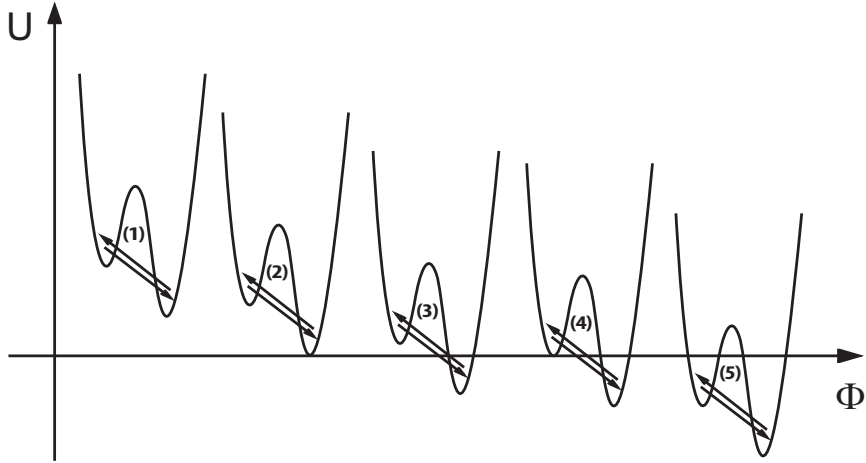}
\end{center}
\caption{\footnotesize{The schematic illustration of the potentials considered in this work. We are interested in the tunneling with five different types of potentials. Arrows show directions of the tunneling transition between vacua. }} \label{fig:fig00}
\end{figure}

We are interested in the tunneling with five different types as shown in Fig.\ \ref{fig:fig00}. Arrows show directions of the tunneling transition between vacua. Type (1) indicates the tunneling between dS-dS vacua. Type (2) for the tunneling between dS-flat vacua. Type (3) for the tunneling between dS-AdS vacua. Type (4) for the tunneling between flat-AdS vacua. Type (5) for the tunneling between AdS-AdS vacua.

After Euclidean rotation we take an $O(4)$ symmetry for both $\Phi$ and the metric $g_{\mu\nu}$, expecting its dominant contribution \cite{cgm}.
The general $O(4)$-symmetric Euclidean metric takes the form
\begin{equation}
ds^2 = d\eta^2 + \rho^2(\eta)[d\chi^2 +
\sin^2\chi(d\theta^2+\sin^2\theta d\phi^2)]\,.  \label{gemetric}
\end{equation}
Then, $\Phi$ and $\rho$ depend only on $\eta$ and the Euclidean field equations for them become
\begin{equation}
\Phi'' + \frac{3\rho'}{\rho}\Phi'=\frac{dU}{d\Phi} \,\,\,\, {\rm
and} \,\,\,\, \rho'' = - \frac{\kappa}{3}\rho (\Phi'^2 +U)\,,
\label{ephi}
\end{equation}
respectively and the Hamiltonian constraint is given by
\begin{equation}
\rho'^2 - 1 -
\frac{\kappa\rho^2}{3}\left(\frac{1}{2}\Phi'^{2}-U\right) = 0\, ,
\label{erho}
\end{equation}
where the prime denotes the derivative with respect to $\eta$.

We now consider vacuum bubble solutions in Einstein gravity. After reviewing true vacuum bubbles and bounce solutions for degenerate vacua, we classify the possible types of false vacuum bubbles. We call `flat' for the geometry with the vanishing cosmological constant, `dS' for the positive cosmological constant, and `AdS' for the negative cosmological constant. We call also `small', `half', and `large' if the portion of the dS space does not exceed half, is equal to half, and does exceed half, respectively. We call `finite' for the geometry with a compact size. According to the terms \cite{ll03} we omit mentioning `infinite' for the non-compact geometry if there is no confusion. We distinguish between half and small dS geometry using the thin-wall approximation despite the fact that the distinctions between half and small dS geometry is not clear among the numerical solutions.

\begin{figure}
\begin{center}
\includegraphics[width=4.8 in]{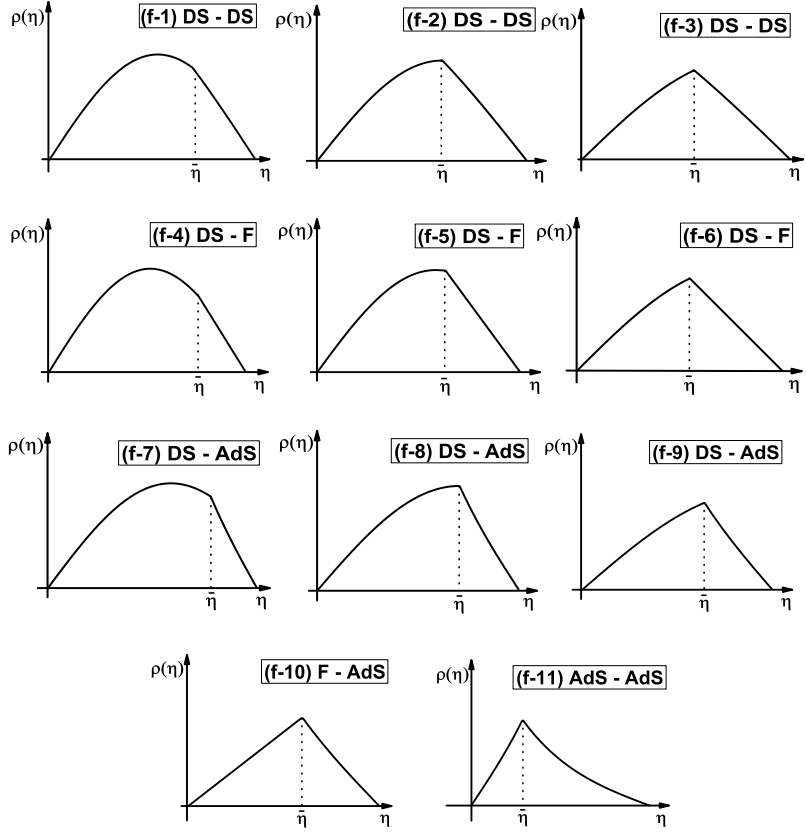}
\end{center}
\caption{\footnotesize{The schematic diagrams for $11$ possible types
of false vacuum bubbles with the compact true vacuum geometry. They are possible as numerical solutions.}} \label{fig:fig01}
\end{figure}

\begin{figure}
\begin{center}
\includegraphics[width=4.8 in]{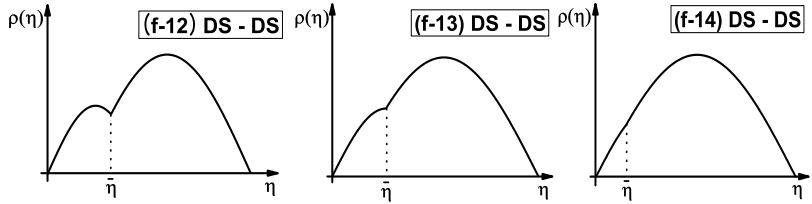}
\end{center}
\caption{\footnotesize{The schematic diagrams for extra $3$ possible types
of matching with the thin-wall approximation. These solutions are not possible.}} \label{fig:fig01-2}
\end{figure}

Firstly, we review true vacuum bubbles. The flat bubble with large dS geometry and the AdS bubble with infinite flat geometry were first investigated in Ref.\ \cite{bnu02}, in which the nucleation rate and the radius of a bubble were obtained by employing the thin-wall approximation. The general cases with an arbitrary vacuum energy, i.e. small dS bubble with large dS geometry, flat bubble with large dS geometry, AdS bubble with large dS geometry, AdS bubble with infinite flat geometry, and AdS bubble with infinite AdS geometry, were investigated in Ref.\ \cite{par02}, in which the general formula for the above cases of the true vacuum bubble was obtained. The spontaneous creation of a membranes \cite{brte}, the decay of metastable state as the conformal vacuum \cite{rusi}, and four different instantons corresponding to true vacuum bubble in a dS geometry was studied in Ref.\ \cite{matu}. The possible types of the true vacuum bubbles within the dS geometry were studied in Ref.\ \cite{ll03}, in which six types of the true vacuum bubble were analyzed in more detail. The case representing the AdS bubble with finite flat geometry \cite{bj05} and the AdS bubble with finite AdS geometry were studied \cite{bfl05}. In Ref.\ \cite{bj05} the authors studied the case of the AdS bubble with large dS geometry. The geometry is compact, which means $\rho$ is zero at $\eta_{max}$. They continuously reduced the false vacuum energy to zero while maintaining $\rho=0$ at $\eta_{max}$. This bounce solution corresponds to a domain wall-like solution. The separation of regions between `finite' flat geometry with $\rho=0$ at $\eta_{max}$ and `infinite' flat geometry with $\rho=\infty$ at $\eta_{max}(=\infty)$ is called Grate Divide. In Ref.\ \cite{bfl05} the authors also studied the AdS bubble with the outside compact geometry, i.e.\ $\rho=0$ at $\eta_{max}$, by lowering the false vacuum energy from positive to negative.

The possible types of true vacuum bubbles are as follows:
(t-1) small dS bubble - large dS geometry, (t-2) small dS bubble - half dS geometry, (t-3) small dS bubble - small dS geometry, (t-4) flat bubble - large dS geometry, (t-5) flat bubble - half dS geometry, (t-6) flat bubble - small dS geometry, (t-7) AdS bubble - large dS geometry, (t-8) AdS bubble - half dS geometry, (t-9) AdS bubble - small dS geometry, (t-10) AdS bubble - infinite flat geometry, (t-11) AdS bubble - infinite AdS geometry, (t-12) AdS bubble - finite flat geometry, (t-13) AdS bubble - finite AdS geometry.

From the cases (t-1) to (t-11), the bubbles correspond to the ordinary CdL-type bubbles. The additional cases are (t-12) and (t-13). For the cases (t-10) and (t-11) the solutions have the infinite geometry. The others have the compact geometry.

Secondly, we briefly review the bounce solutions for degenerate vacua \cite{hw10, lllo10, bll90, lllo67}. The possible cases are (d-1) dS - dS \cite{hw10, lllo10}, (d-2) flat - flat, and (d-3) AdS - AdS \cite{lllo10}. All cases have the compact geometry with $Z_2$ symmetry. The mechanism for making the domain wall or braneworld-like object is different from the ordinary formation mechanism of the domain wall because our solutions are instanton (bounce) solutions rather than soliton solutions \cite{lllo10}. In other words, our solutions can be interpreted as solutions corresponding to the domain wall by an instanton-induced theory rather than a soliton-induced theory. The oscillating solutions with $Z_2$ symmetry \cite{llwc1} could be interpreted as a mechanism providing nucleation of the thick wall for topological inflation \cite{avil04, avil08}.

We now investigate false vacuum bubbles. In general, false vacuum bubbles can occur only if gravity is taken into account. The false one with the non-compact geometry is not allowed in Einstein gravity. This can be seen in Eqs.\ (\ref{gemetric}), (\ref{ephi}), and (\ref{erho}). In other words, those with non-compact geometry should correspond to that of the black hole, which can not be matched to the bubble geometry with $O(4)$ symmetry. Thus we consider the false vacuum bubbles only with compact geometry in Einstein gravity. Some of them within dS geometry were studied in Refs.\ \cite{kw04} and \cite{ll03}.

The schematic diagrams for 14 possible types of false vacuum bubbles with the compact true vacuum geometry or matching with the thin-wall approximation are illustrated in Figs.\ \ref{fig:fig01} and \ref{fig:fig01-2}. The figures of the first row (f-1) - (f-3) represent the dS interior and dS exterior geometry, those of the second row (f-4) - (f-6) represent the dS interior and flat exterior geometry, those of the third row (f-7) - (f-9) represent the dS interior and AdS exterior geometry, and those of the fourth row (f-10) and (f-11) represent the flat interior, AdS interior and AdS exterior geometry, respectively. The cases in Fig.\ \ref{fig:fig01-2} do not have a stationary point in the action. Therefore, these solutions are not possible.

We classify the possible types of false vacuum bubbles as follows (See, Fig.\ \ref{fig:fig01}):
(f-1) large dS bubble - small dS geometry, (f-2) half dS bubble - small dS geometry, (f-3) small dS bubble - small dS geometry, (f-4) large dS bubble - finite flat geometry, (f-5) half dS bubble - finite flat geometry, (f-6) small dS bubble - finite flat geometry, (f-7) large dS bubble - finite AdS geometry, (f-8) half dS bubble - finite AdS geometry, (f-9) small dS bubble - finite AdS geometry, (f-10) flat bubble - finite AdS geometry, (f-11) AdS bubble - finite AdS geometry \cite{lllo67}.

All cases have the compact geometry. Among them in Fig.\ \ref{fig:fig01}, only $11$ types of matching can be obtained as the numerical solutions (Fig.\ \ref{fig:fig01}). From (f-1) to (f-9), the false vacuum bubbles have the positive vacuum energy, i.e.\ the cases represent the nucleation of dS space. If we compare the false vacuum bubbles with the true ones, we can make a pair between the false and true one. The pairs consist of ((f-1) and (t-1)), ((f-2) and (t-2)), ((f-3) and (t-3)), ((f-4) and (t-4)), ((f-5) and (t-5)), ((f-6) and (t-6)), ((f-7) and (t-7)), ((f-8) and (t-8)), ((f-9) and (t-9)), ((f-10) and (t-12)), and ((f-11) and (t-13)). If we employ the transform $\eta \rightarrow \eta_{max}-\eta$, the bounce solutions for a false vacuum bubble could be expected. The important thing is to show the existence of the false vacuum bubbles as the numerical solutions and obtain the nucleation probability. The implications of the solutions could be also important when the solutions could be applied to the history of the very early universe.

In what follows, we suppose that the background space initially resides in the homogeneous true vacuum, $\Phi_T$. Due to the asymmetry of the potential, the probability for $\Phi_T$ selected as the initial background state is slightly larger than the probability for $\Phi_F$. The probabilities for both $\Phi_T$ and $\Phi_F$ will be same in the case of the vanishing $\epsilon$, i.e.\ the case under the potential with degenerate minima. Although the higher probability for $\Phi_T$, there is a significant disparity between two situations. The tunneling from $\Phi_F$ to $\Phi_T$, down-tunneling, is quite natural. However, the tunneling from $\Phi_T$ to $\Phi_F$, up-tunneling, is not quite natural. In the absence of gravity, the event may not occur. Our work is devoted to the event in the presence of Einstein gravity. The strategy for obtaining our solutions is straightforward. We impose the boundary conditions in accord with the solution what we want to obtain. After finding the numerical solutions yielding above equations, we will calculate the radius and the nucleation rate of a vacuum bubble.

We now impose the boundary conditions to solve Eqs.\ (\ref{ephi}) and (\ref{erho}) in accord with the new solutions with compact geometry. For this purpose, the values of the field $\rho$ and derivatives of the field $\Phi$ are imposed at both ends of the evolution parameter, i.e.\
$\eta=0$ and $\eta=\eta_{max}$ as follows:
\begin{equation}
\rho|_{\eta=0}=0 ,
\,\,\,\, \rho|_{\eta =\eta_{max}} =0, \,\,\,\,
\frac{d\Phi}{d\eta}\Big|_{\eta=0}=0, \,\,\,\, {\rm and}\,\,\,\,
\frac{d\Phi}{d\eta}\Big|_{\eta =\eta_{max}} = 0, \label{ourbc-2}
\end{equation}
where $\eta_{max}$ is the maximum value of $\eta$ and is a finite value in this work. The first two conditions are for the background space. The last two conditions are for the scalar field. The first condition is to obtain a geodesically complete spacetime. The second condition is to obtain a compact geometry. For dS background the second condition is natural. The third and fourth conditions guarantee that the solutions are regular at both ends as can be seen from the first equation in Eq.\ (\ref{ephi}). We now mention one superfluous condition on the initial value of $\Phi|_{\eta=0}=\Phi_o$, which is not determined. One should find the initial value of $\Phi$ using the undershoot-overshoot procedure. Some initial $\Phi_o$ will give the overshooting, in which the value of $\Phi$ at late $\eta$ value will go beyond $\Phi_T$ or $\Phi_F$ with a nonvanishing velocity. Some other initial value $\Phi_o$ will give the undershoot, in which the value of $\Phi$ at late $\eta$ does not climb all the way up to $\Phi_T$ or $\Phi_F$. Thus, the value $\Phi_o$ should exist within an intermediate position between the undershoot and overshoot \cite{col002}. Then, the solution yields a maximum of the action rather than a minimum, i.e.\ it gives the saddle point of the action. This implies the existence of a negative mode as a bounce solution. The negative mode is caused by the variation for the behavior of the magnitude of the action depending on the magnitude of the radius of a bounce solution \cite{cacol, cacol01, cacol02, cacol03, cacol04}. We employ these boundary conditions for our bounce solutions.

We note that the condition for new solutions is to have the existence of $\rho'=0$. In other words, it can be given by
\begin{equation}
U (\Phi_{top})> 0 \,, \label{condpoten}
\end{equation}
where $U(\Phi_{top})$ indicates the local maximum point of the potential $U (\Phi)$. If the parameter $\epsilon$ is equal to zero, then the condition is reduced to the follows: $U_o > -\mu^4/8\lambda$ \cite{lllo10}.

\section{The nucleation of a false vacuum bubble \label{sec3}}

We solve the coupled equations for the gravity and the scalar field simultaneously satisfying with boundary conditions Eq.\ (\ref{ourbc-2}). After finding the numerical solutions we evaluate the radius and the nucleation rate of a vacuum bubble using the thin-wall approximation.

\subsection{Numerical solutions \label{sec3-1}}

We rewrite the equations in terms of dimensionless variables as in Ref.\ \cite{lllo10}. We employ the shooting method using the adaptive step size Runge-Kutta as done in Ref.\ \cite{ptvf}. For this procedure we choose the initial values of $\tilde{\Phi}(\tilde{\eta}_{initial})$, $\tilde{\Phi}'(\tilde{\eta}_{initial})$, $\tilde{\rho}(\tilde{\eta}_{initial})$, and $\tilde{\rho}'(\tilde{\eta}_{initial})$ at $\tilde{\eta}=\tilde{\eta}_{initial}$ as follows:
\begin{eqnarray}
\tilde{\Phi}(\tilde{\eta}_{initial}) &\sim& \tilde{\Phi}_{o} +
\frac{\varepsilon^2}{16}[\tilde{\Phi}_{o}\left( \tilde{\Phi}_{o}^2-1
\right) -\epsilon] +
\frac{\varepsilon^{3}}{384}[\tilde{\Phi}_{o}\left( \tilde{\Phi}_{o}^2-1
\right) -\epsilon]\left( 3\tilde{\Phi}_{o}^{2} - 1 \right) + \cdot\cdot\cdot,  \nonumber \\
\tilde{\Phi}'(\tilde{\eta}_{initial}) &\sim& \frac{\varepsilon}{8}[
\tilde{\Phi}_{o} \left( \tilde{\Phi}_{o}^{2} - 1 \right) -\epsilon]+
\frac{\varepsilon^{2}}{128}[\tilde{\Phi}_{o}\left( \tilde{\Phi}_{o}^2-1
\right) -\epsilon]\left( 3\tilde{\Phi}_{o}^2
- 1 \right) + \cdot\cdot\cdot ,  \label{nbcon2} \\
\tilde{\rho}(\tilde{\eta}_{initial}) &\sim& \varepsilon + \cdot\cdot\cdot,  \nonumber \\
\tilde{\rho}'(\tilde{\eta}_{initial}) &\sim& 1 + \cdot\cdot\cdot , \nonumber
\end{eqnarray}
where $\tilde{\eta}_{initial} = 0+\varepsilon$ and $\varepsilon \ll 1$. In numerical calculation, we take $\tilde{\eta}_{initial} = \varepsilon$ instead of $0$ to obtain a regular evolution as the first step, as one can check the smoothness of each term in Eq.\ (\ref{ephi}). The functions $\tilde{\Phi}(\tilde{\eta}_{initial})$, $\tilde{\Phi}'(\tilde{\eta}_{initial})$, $\tilde{\rho}(\tilde{\eta}_{initial})$, and $\tilde{\rho}'(\tilde{\eta}_{initial})$ are smooth, so that one can employ the Taylor expansions of these functions around the initial values as above. After finding the initial value $\tilde{\Phi}_{o}$ numerically, the other conditions are given by Eqs.\ (\ref{nbcon2}).

In order to ensure the regular solution at $\tilde{\eta} = \tilde{\eta}_{max}$ in Eq.\ (\ref{ephi}), we demand the conditions $d\tilde{\Phi}/{d\tilde{\eta}} \rightarrow 0$ and $\tilde{\rho}\rightarrow 0$ as $\tilde{\eta}\rightarrow \tilde{\eta}_{max}$. In this work, the exact value of $\tilde{\eta}_{max}$ is not known.

\begin{figure}[t]
\begin{center}
\includegraphics[width=5. in]{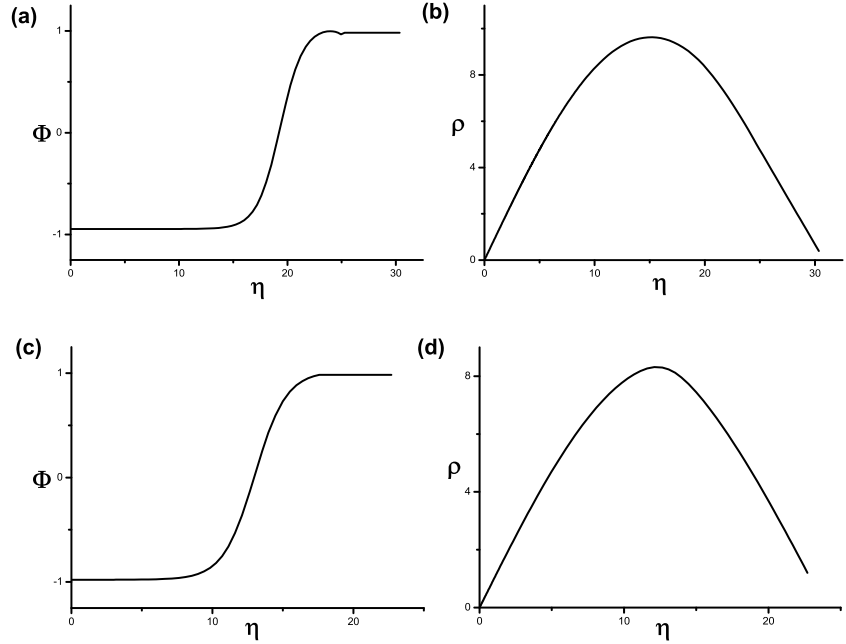}
\end{center}
\caption{\footnotesize{dS-dS cases. (a) and (b) correspond to the case (f-1) in Fig.\ \ref{fig:fig01}, i.e.\ large dS bubble - small dS geometry, while (c) and (d) correspond to the case (f-3) in Fig.\ \ref{fig:fig01}, i.e.\ small dS bubble - small dS geometry.}} \label{fig:fig02}
\end{figure}

Figure \ref{fig:fig02} shows the numerical solutions for $\Phi$ and $\rho$. Figures \ref{fig:fig02}(a) and \ref{fig:fig02}(b) correspond to the case (f-1) in Fig.\ \ref{fig:fig01}, i.e.\ large dS bubble - small dS geometry, in which the position of the wall is over the location of $\tilde{\eta}$ corresponding to the maximum value of $\tilde{\rho}$ for dS bubble. While figures \ref{fig:fig02}(c) and \ref{fig:fig02}(d) correspond to the case (f-3) in Fig.\ \ref{fig:fig01}, i.e.\ small dS bubble - small dS geometry, in which the width of the wall includes the location of $\tilde{\rho}_{max}$. The $\tilde{\rho}_{max}$ does not reach at the maximum value of $\tilde{\rho}$ for the inside dS bubble or dS geometry. We take $\tilde{\epsilon} = 0.1$, $\tilde{\kappa} = 0.05$, and $\tilde{U_o} = 0.55$ for top figures, while $\tilde{\epsilon} = 0.04$, $\tilde{\kappa} = 0.30$, and $\tilde{U_o} = 0.01$ for bottom figures.

\begin{figure}[t]
\begin{center}
\includegraphics[width=5. in]{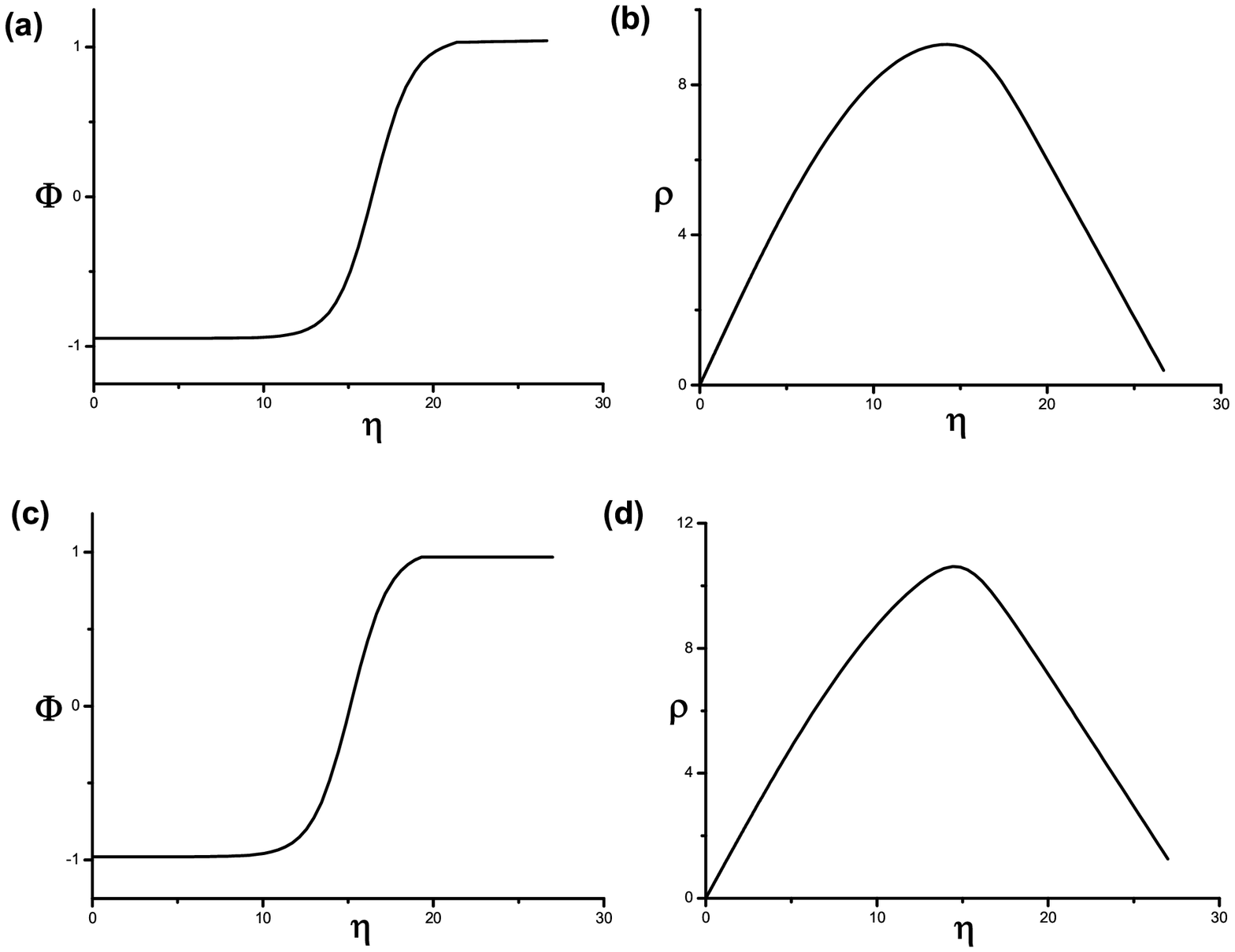}
\end{center}
\caption{\footnotesize{ds-flat cases. (a) and (b) correspond to the case (f-4) in Fig.\ \ref{fig:fig01}, i.e.\ large dS bubble - finite flat geometry, while (c) and (d) correspond to the case (f-6) in Fig.\ \ref{fig:fig01}, i.e.\ small dS bubble - finite flat  geometry.}} \label{fig:fig03}
\end{figure}

Figure \ref{fig:fig03} shows the numerical solutions for $\Phi$ and $\rho$. Figures \ref{fig:fig03}(a) and \ref{fig:fig03}(b) correspond to the case (f-4) in Fig.\ \ref{fig:fig01}, i.e.\ large dS bubble - finite flat geometry, while figures \ref{fig:fig03}(c) and \ref{fig:fig03}(d) correspond to the case (f-6) in Fig.\ \ref{fig:fig01}, i.e.\ small dS bubble - finite flat geometry. We take $\tilde{\epsilon} = 0.1$, $\tilde{\kappa} = 0.37$, and $\tilde{U_o} = 0.0012$ for top figures. $\tilde{\epsilon} = 0.04$, $\tilde{\kappa} = 0.5$, and $\tilde{U_o}= 0.0077$ for bottom figures.

\begin{figure}[t]
\begin{center}
\includegraphics[width=5. in]{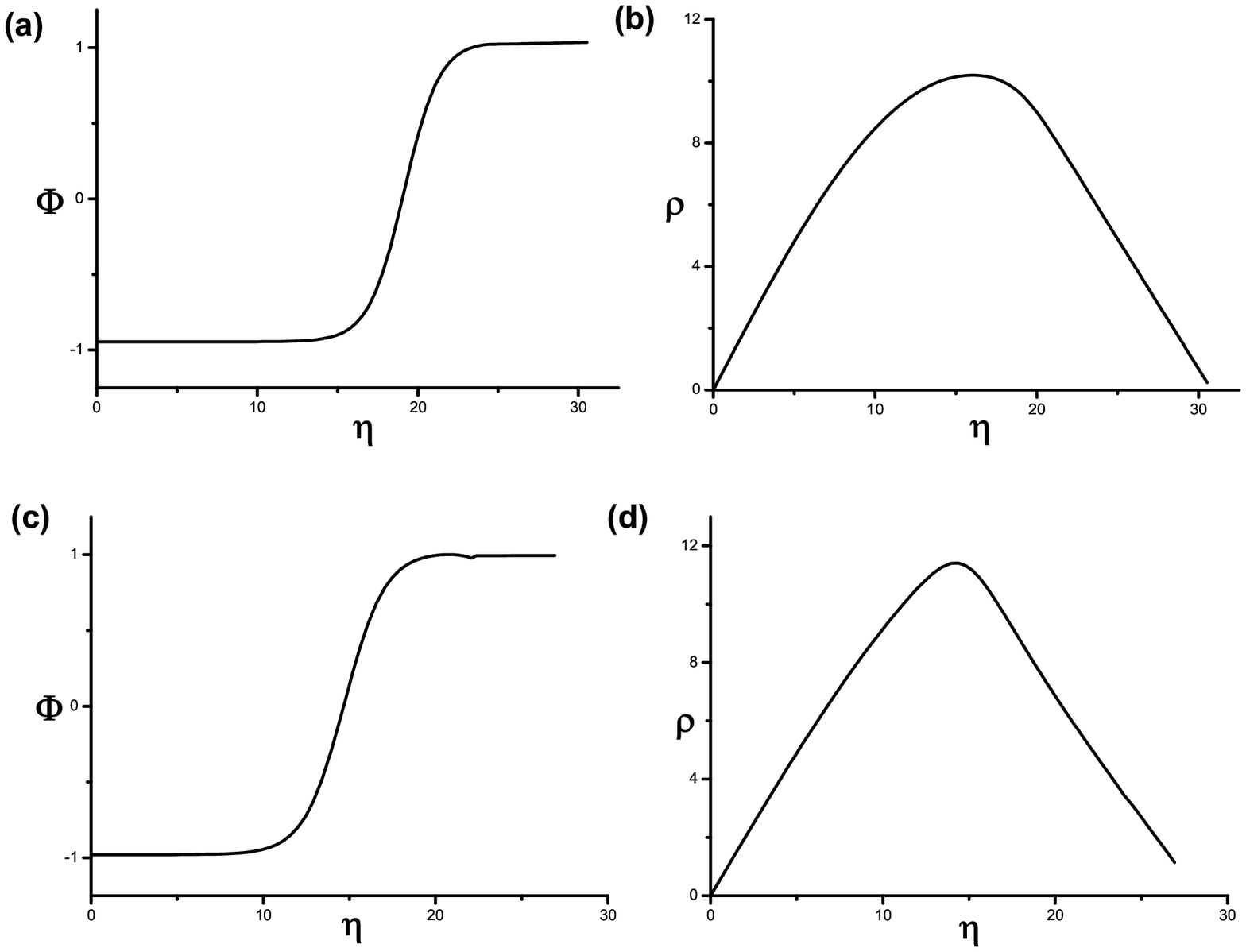}
\end{center}
\caption{\footnotesize{dS-AdS cases. (a) and (b) correspond to the case (f-7) in Fig.\ \ref{fig:fig01}, i.e.\ large dS bubble - finite AdS geometry, while (c) and (d) correspond to the case (f-9) in Fig.\ \ref{fig:fig01}, i.e.\ small dS bubble - finite AdS geometry.}} \label{fig:fig04}
\end{figure}

Figure \ref{fig:fig04} shows the numerical solutions for $\Phi$ and $\rho$. Figures \ref{fig:fig04}(a) and \ref{fig:fig04}(b) correspond to the case (f-7) in Fig.\ \ref{fig:fig01}, i.e.\ large dS bubble - finite AdS geometry, while figures \ref{fig:fig04}(c) and \ref{fig:fig04}(d) correspond to the case (f-9) in Fig.\ \ref{fig:fig01}, i.e.\ small dS bubble - finite AdS geometry. We take $\tilde{\epsilon} = 0.10$, $\tilde{\kappa} = 0.3$, and
$\tilde{U_o} = -0.001$ for top figures. $\tilde{\epsilon} = 0.04$, $\tilde{\kappa} = 0.8$, and $\tilde{U_o} = -0.02$ for bottom figures.

\begin{figure}[t]
\begin{center}
\includegraphics[width=5. in]{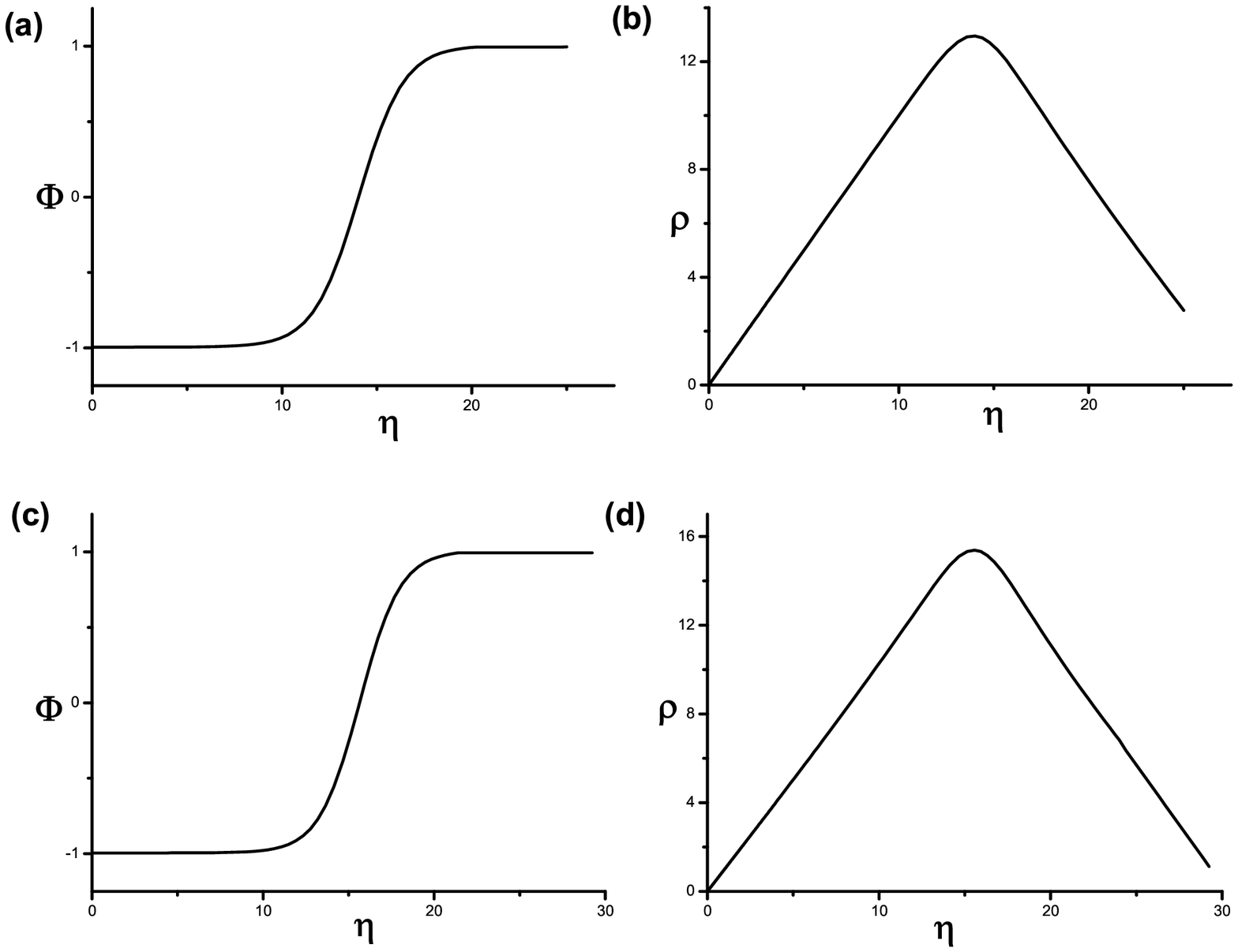}
\end{center}
\caption{\footnotesize{Flat-AdS and AdS-AdS cases. (a) and (b) correspond to the case (f-10) in Fig.\ \ref{fig:fig01}, i.e.\ flat bubble - finite AdS geometry, while figures (c) and (d) correspond to the case (f-11) in Fig.\ \ref{fig:fig01}, i.e.\ AdS bubble - finite AdS geometry.}} \label{fig:fig05}
\end{figure}

Figure \ref{fig:fig05} shows the numerical solutions for $\Phi$ and $\rho$. Figures \ref{fig:fig05}(a) and \ref{fig:fig05}(b) correspond to the case (f-10) in Fig.\ \ref{fig:fig01}, i.e.\ flat bubble - finite AdS geometry, while figures \ref{fig:fig05}(c) and \ref{fig:fig05}(d) correspond to the case (f-11) in Fig.\ \ref{fig:fig01}, i.e.\ AdS bubble - finite AdS geometry. We take
$\tilde{\epsilon} = 0.01$, $\tilde{\kappa} = 0.95$, and $\tilde{U_o} = -0.01$ for flat-AdS case. $\tilde{\epsilon} = 0.01$, $\tilde{\kappa} = 0.95$, and $\tilde{U_o} = -0.015$ for AdS-AdS case.

\subsection{Nucleation rate \label{sec3-2}}

We now evaluate the nucleation rate $e^{-B}$. The exponent $B$ is the difference between the action of the bubble solution and that of the background geometry. The Euclidean action is given by
\begin{equation}
S_E= \int_{\mathcal M} \sqrt{g_E} d^4 x \left[- \frac{R}{2\kappa}
+\frac{1}{2}{\nabla^\alpha}\Phi {\nabla_\alpha}\Phi
+U(\Phi)\right] - \oint_{\partial \mathcal M} \sqrt{h_E} d^3 x
\frac{K-K_o}{\kappa}\, . \label{eu-action}
\end{equation}
We consider the action for the background space. The Euclidean dS space is a compact geometry, so that there is no YGH boundary term. The boundary term can contribute to the action for flat Minkowski and AdS space. The YGH boundary term with the extrinsic curvature $K$ has the contribution from both the vacuum energy and the shape of the geometry at the boundary. The YGH boundary term with the subtraction $K_o$ can be introduced to ensure the action vanishing identically for flat space, i.e., the term cancels out the contribution from the shape of the geometry at the boundary. As a result, it gives $S_E(\rm{flat}) =0$. The Euclidean action for AdS space has the contribution of $+\infty$ from the bulk part and $-\infty$ from the boundary term. However, the contribution from the boundary term is greater than that from the bulk part. Thus, it gives $S_E(\rm{AdS}) =-\infty$. This property could be renormalized by introducing a specific counter term.

We now consider the action for both the solution and the background space. We note that both false vacuum bubbles and the background are the solutions of equations of the motion with $O(4)$ symmetry. For $O(4)$-symmetric solution, the bulk part in Eq.\ (\ref{eu-action}) can be divided into two parts after integration by parts, which makes the action as
\begin{equation}
S_E = 4\pi^2 \int^{\eta_{max}}_{0} d\eta \left[\rho^3 U - \frac{3\rho}{\kappa} \right] + \frac{6\pi^2}{\kappa}(\rho^2\rho')|_{\eta_{max}} -\frac{2\pi^2 \rho^3}{\kappa}(K-K_o)|_{\eta_{max}}  \,, \label{fe-action}
\end{equation}
where $K=\frac{3\rho'}{\rho}|_{\eta_{max}}$ and $K_o=\frac{3}{\eta}|_{\eta_{max}}$. The surface term from the parts integration, $\frac{6\pi^2}{\kappa}(\rho^2\rho')|_{\eta_{max}}$, and the YGH boundary term are harmless if one is interested in the action difference between the solution and the background that agree at asymptotics \cite{bnu02}. Actually, the surface term is exactly canceled by the boundary term, $-\frac{2\pi^2 \rho^3}{\kappa}K|_{\eta_{max}}$ \cite{wein007}. The dS space with compact geometry has neither the surface term nor YGH boundary term. This is true for our solutions with compact geometries, while the backgrounds for flat and AdS space have both two terms. We perform the analytic computation carefully with this point.

We evaluate the nucleation rate of a false vacuum bubble. The fields vary continuously between one and the other vacuum state as shown in numerical solutions. If the coefficient $\epsilon$ in Eq.\ (\ref{poten}) is small in comparison with all other parameters of the model, we can make use of the thin-wall approximation scheme to evaluate $B$. The validity of that in the case of a true vacuum bubble has been examined \cite{samuel, samuel01}. We assume the thin-wall approximation is still valid in this work. Analogous to the case in the absence of gravity, the Euclidean action could be divided into three parts: $B= B_{in} + B_{wall} + B_{out}$.

The contribution of the wall is given by $B_{wall}=2\pi^2 \bar\rho^3S_o$, in which $2\pi^2 \bar\rho^3$ is treated as a constant inside the wall. The surface density or the tension of the wall $S_o( = \frac{2\mu^3}{3\lambda})$ is a positive constant \cite{col002, orlan}. The $\epsilon$-dependent term in Eq.\ (\ref{poten}) and the damping term in Eq.\ (\ref{ephi}) were neglected in the leading order \cite{col002}. We note that the $\cal{O}(\epsilon)$ of contribution to $B_{wall}$ by $\epsilon$ has different signs, i.e.\ $+$ contribution for up-tunneling and $-$ contribution for down-tunneling.

We now compute the contribution from the inside of the wall $B_{in} (=S^F_{in}-S^T_{in})$, where the $S^F_{in}$ corresponds to the false vacuum configuration inside of the wall, while $S^T_{in}$ corresponds to the true vacuum configuration of the background inside of the wall. The expression will depend on the size of the bubble and the background geometry. As pointed out in Ref.\ \cite{ll03}, if the size of dS bubble  is larger than half of its dS size,  the integration range of the false vacuum region in $B_{in}$  should be divided into two parts. Likewise, if the size of dS background is larger than that, the range should be also  divided into two parts as for the case (f-1) in Fig.\ \ref{fig:fig01}. For example, the formula of $B_{in}$ for the case (f-1) can be described as
\begin{eqnarray}
B_{in} &=& S^F_{in}-S^T_{in}= 4\pi^2 \left[ \int^{\rho_{max}}_{0} d\rho \frac{\rho^3U_F - \frac{3\rho}{\kappa}}{(1-\frac{\kappa}{3}\rho^2
U_F)^{1/2}} -  \int^{\bar\rho}_{\rho_{max}} d\rho \frac{\rho^3U_F - \frac{3\rho}{\kappa}}{(1-\frac{\kappa}{3}\rho^2
U_F)^{1/2}} \right] \nonumber \\
&-& 4\pi^2 \left[ \int^{\rho_{max}}_{0} d\rho \frac{\rho^3U_T - \frac{3\rho}{\kappa}}{(1-\frac{\kappa}{3}\rho^2
U_T)^{1/2}} -  \int^{\bar\rho}_{\rho_{max}} d\rho \frac{\rho^3U_T - \frac{3\rho}{\kappa}}{(1-\frac{\kappa}{3}\rho^2
U_T)^{1/2}} \right]\nonumber \\
&=& \frac{12\pi^2}{\kappa^2}\left[\frac{-1-(1-\frac{\kappa}{3}\bar\rho^2 U_F)^{3/2}}{U_F} -
\frac{-1-(1-\frac{\kappa}{3}\bar\rho^2 U_T)^{3/2}}{U_T}
 \right]\, ,
\label{lee-wein}
\end{eqnarray}
where $\rho_{max}(F/T)=\sqrt{3/\kappa U_{F/T}}$ for dS geometry and we use the relation
\begin{equation}
d\rho =\pm d\eta \left[1-\frac{\kappa\rho^2U}{3}\right]^{1/2},
\label{lee001-inw}
\end{equation}
where $+$ is for $0 \le \eta < \eta_{max}/2$, $0$ is for
$\eta=\eta_{max}/2$, and $-$ is for $\eta_{max}/2 < \eta \le
\eta_{max}$, respectively. The formula of $B_{in}$ for the false vacuum bubbles can be summarized as
\begin{equation}
B_{in} =\frac{12\pi^2}{\kappa^2}\left[\frac{-1\pm(1-\frac{\kappa}{3}\bar\rho^2 U_F)^{3/2}}{U_F} -
\frac{-1\pm(1-\frac{\kappa}{3}\bar\rho^2 U_T)^{3/2}}{U_T}
 \right] \, . \label{b_in}
\end{equation}
Here the $+$ sign in front of $(1-\frac{\kappa}{3}\bar\rho^2 U_F)^{3/2}$ is for the small bubble, (f-3), (f-6), (f-9), (f-10), and (f-11), while the $-$ sign is for the large dS bubble, (f-1), (f-4), and (f-7). For the half dS bubble, $(1-\frac{\kappa}{3}\bar\rho^2 U_F)^{3/2}$ is vanishing. The $+$ sign in front of $(1-\frac{\kappa}{3}\bar\rho^2 U_T)^{3/2}$ is for the flat or AdS background, (f-4), (f-5), (f-6), (f-7), (f-8), (f-9), (f-10), and (f-11), while the $-$ sign is for the dS background, (f-1), (f-2), and (f-3). As a mnemonic, the $+$ signature is for ``small", while the $-$ signature is for ``large". This holds true also for $B_{in}$ of the true vacuum bubble solutions, which can be obtained by switching $U_T \leftrightarrow U_F$ and by keeping only the $+$ sign of the first term changed by $(1-\frac{\kappa}{3}\bar\rho^2 U_T)^{3/2}$. This is because the true vacuum bubble is ``small" as can be seen in Fig.\ $1$ in Ref.\ \cite{ll03}. The true vacuum bubbles in Ref.\ \cite{par02} correspond to taking positive signs in both terms, i.e.\ the small true vacuum bubbles and the small background in the inside part.

We note that the contribution to the action from the inside of the wall is negative for both the true vacuum, $S^T_{in} < 0$, and the false vacuum configuration, $S^F_{in} < 0$, while $S^T_{in} < S^F_{in}$. Therefore $B_{in}=S^F_{in}-S^T_{in} > 0$ for the false vacuum bubbles. This is opposite sign to that of the CdL-type true vacuum bubbles, $B^{CD}_{in}=S^T_{in}-S^F_{in} < 0$.

We now consider the contribution from the outside of the wall $B_{out}$. For the CdL-type true vacuum bubble, there is no contribution to $B$, i.e., $B_{out}=0$. For the false vacuum bubbles, this is true only for dS space. This is due to the same contribution from the bubble solution and the background space. For the flat and AdS space the evaluation of $B_{out}$ is more subtle. The integration by parts should be carried out carefully. Our solutions with compact geometries have neither the surface term nor YGH boundary term. However, the backgrounds for flat and AdS space have both two terms in Eq.\ (\ref{fe-action}), which is located at $\eta_{max}$. The size of the background space will go to infinity as $\tilde\rho^i_{max}(\eta_{max})$ goes to infinity. Therefore, the contribution from the outside part in both flat and AdS geometry is evaluated to be
\begin{eqnarray}
B_{out} &=& 4\pi^2 \left[ \int^0_{\bar{\rho}} (-d\rho) \frac{(\rho^3U_T-\frac{3\rho}{\kappa})}{[1-\frac{\kappa}{3}\rho^2U_T]^{1/2}} - \int^{\tilde\rho^{i}_{max}}_{\bar{\rho}} d\rho \frac{(\rho^3U_T-\frac{3\rho}{\kappa})}{[1-\frac{\kappa}{3}\rho^2U_T]^{1/2}} \right] \nonumber \\
&+&B_{sur} + \frac{2\pi^2\rho^3}{\kappa}(K-K_o)|_{\eta_{max}},
\end{eqnarray}
where the first term in the right-hand-side is from the outside of a solution and the second term is from the background space. The surface term $B_{sur} = - \frac{6\pi^2}{\kappa}(\rho^2\rho')|_{\eta_{max}}$ and the YGH boundary term $\frac{2\pi^2\rho^3}{\kappa}K|_{\eta_{max}}$ coming only from the background space are canceled out. Then the remaining terms in the action become
\begin{equation}
B_{out}=\frac{12\pi^2}{\kappa^2
U_T}\left[2\left(1-\frac{\kappa}{3}\bar\rho^2 U_T \right)^{3/2}
\right] +B_{feff}\,, \label{b_out}
\end{equation}
where
\begin{equation}
B_{feff}=- \frac{12\pi^2}{\kappa^2 U_T} \left[1 +
\left(1- \frac{\kappa}{3}\tilde\rho^{i2}_{max} U_T \right)^{3/2}
\right] - \frac{2\pi^2\rho^3}{\kappa}K_o|_{\eta_{max}}\,. \label{b-feff}
\end{equation}

We briefly summarize the results of $B_{feff}$ in Eq.\ (\ref{b-feff}). For the dS bubble with the dS background space, which corresponds to the cases (f-1), (f-2) and (f-3), $B_{feff}$ (and $B_{out}$) is vanishing. This is because both the dS bubble solution and the dS background space have neither the surface term nor the YGH boundary term. For the flat and AdS background space $B_{feff}$ is not vanishing. For the flat background space, the term $\frac{6\pi^2}{\kappa}\tilde\rho^{i2}_{max}$ is canceled by the subtraction term $- \frac{2\pi^2\rho^3}{\kappa}K_o|_{\eta_{max}}$ in Eq.\ (\ref{b-feff}). The term $-\frac{24\pi^2}{\kappa^2 U_T}$ in $B_{feff}$ is canceled by the first term in the right-hand-side in Eq.\ (\ref{b_out}). Thus the final contribution is $B_{out}=-\frac{12\pi^2}{\kappa}\bar{\rho}^2$. For the AdS background, $B_{feff}=\frac{4\pi^2}{\sqrt\kappa}\sqrt{\frac{|U_T|}{3}}\tilde\rho^{i3}_{max} +{\cal O}(\tilde\rho^{i3}_{max}/\eta_{max})$. As the size of the AdS background space goes to infinity which means $\eta_{max} \rightarrow \infty$, then $B_{feff} \rightarrow + \infty$, hence $B_{out} \rightarrow + \infty$.

We now evaluate the nucleation rate, i.e.\ $B=B_{in} + B_{wall} + B_{out}$. The nucleation rate is given by
\begin{equation}
B=\frac{12\pi^2}{\kappa^2}\left[\frac{-1\pm(1-\frac{\kappa}{3}\bar\rho^2 U_F)^{3/2}}{U_F} -
\frac{-1-(1-\frac{\kappa}{3}\bar\rho^2 U_T)^{3/2}}{U_T}
 \right] + 2\pi^2 \bar\rho^3S_o +B_{feff}\, . \label{befo-gene-formu}
\end{equation}
Here the $+/-$ sign in front of $(1-\frac{\kappa}{3}\bar\rho^2 U_F)^{3/2}$ is for the small/large bubble. We note that the sign in front of $(1-\frac{\kappa}{3}\bar\rho^2 U_T)^{3/2}$ is $-$ for all cases. This sign is determined only by $B_{in}$ for the dS background, which is negative as explained below Eq.\ (\ref{b_in}). For the flat and AdS background space the extra contribution from $B_{out}$ changes the $+$ sign from $B_{in}$ into $-$.

The size $\bar\rho$ of the bubble is determined by extremizing $B$. The general formula for $\bar{\rho}$ is evaluated to be
\begin{equation}
\bar\rho^2 = \frac{\bar\rho_o^2}{D}\, , \label{gene-formu-rho}
\end{equation}
where $\bar{\rho}_o =3S_o/\epsilon$ is the radius of the true vacuum bubble in the absence of gravity. $D=\left[1+2(\frac{\bar\rho_o}{2\lambda_1})^2+(\frac{\bar\rho_o}{2\lambda_2})^4
\right]$, $\lambda_1^2=[3/\kappa(U_F+U_T)]$, and $\lambda_2^2=[3/\kappa(U_F-U_T)]$.
The term $B_{feff}$ does not affect the determination of $\bar{\rho}$. The form of $\bar{\rho}$ is nontrivially the same as those obtained in Refs.\ \cite{par02, ll03}. This can be seen from the observation that the expression $B$ for the false vacuum bubble in Eq.\ (\ref{befo-gene-formu}) and the true vacuum bubble in Refs.\ \cite{par02, ll03} are the same up to the $\bar{\rho}$-dependent terms. The differences are the sign in front of $1/U_{T/F}$ and the existence of the $B_{feff}$, which is irrelevant to getting $\bar{\rho}$. In more detail, the sign of $(1-\frac{\kappa}{3}\bar\rho^2 U_F)^{3/2}$ is positive in Ref.\ \cite{par02}, while both are allowed in Ref.\ \cite{ll03}. There exist two bubbles with the same radius formula in Euclidean dS space of the topology $S^4$. One corresponds to the large bubble, while the other corresponds to the small bubble. For the large one, $U_F-U_T=\epsilon > \frac{3\kappa S_o^2}{4}$, while for the small one, $\epsilon < \frac{3\kappa S_o^2}{4}$. However, the numerical values of the radii are different. The large bubble occur with the small numerical value $\tilde{\kappa}(=\frac{\mu^2}{\lambda}\kappa)$, while the small bubble occur with the large numerical value $\tilde{\kappa}$. In numerical work, $\tilde{\eta}_{max}$ becomes smaller as  $\tilde{\kappa}$ becomes larger.

After plugging $\bar{\rho}$ into Eq.\ (\ref{befo-gene-formu}) the general formula for $B$ is evaluated to be
\begin{equation}
B=\frac{2B_o[\{1+(\frac{\bar{\rho}_o}{2\lambda_1})^2\} + D^{1/2}
]}{[(\frac{\bar{\rho}_o}{2\lambda_2})^4
\{(\frac{\lambda_2}{\lambda_1})^4 -1 \}D^{1/2}]} +B_{feff}\, , \label{gene-formu}
\end{equation}
where $B_o=27\pi^2S_o^4/2\epsilon^3$ is the nucleation rate of a true vacuum bubble in the absence of gravity and $B_{feff}$ is nonvanishing for flat and AdS background space. This result for dS background, where $B_{feff}$ is vanishing, is already derived in Ref.\ \cite{ll03}. We note that the formula for the false vacuum bubble has the $+$ sign in front of $D^{1/2}$ in the numerator \cite{ll03}, while the formula $B$ for the true vacuum bubbles has the $-$ sign in front of $D^{1/2}$ in the numerator \cite{par02, ll03}.

We nontrivially obtained the same form for all false vacuum bubbles by using the following conditions. For large dS bubbles (corresponding to the cases (f-1), (f-4), and (f-7)), which occur when $\epsilon >3\kappa S^2_o/4$ or equivalently when $1> (\frac{\bar{\rho}_o}{2\lambda_2})^2$, we used the relation $(1-\frac{\kappa}{3}U_F\bar\rho^2)^{3/2}
=\frac{[1-(\frac{\bar{\rho}_o}{2\lambda_2})^2]^3}{D^{3/2}}$. For small bubbles, (f-3), (f-6), (f-9), (f-10), and (f-11), which occur when $\epsilon <3\kappa S^2_o/4$ or $(\frac{\bar{\rho}_o}{2\lambda_2})^2 > 1$, we used the relation $(1-\frac{\kappa}{3}U_F\bar\rho^2)^{3/2}=\frac{[(\frac{\bar{\rho}_o}{2\lambda_2})^2-1]^3}{D^{3/2}}$.

We explain $B$ more detail for each case. We first explain the $B$ with the dS background for the cases (f-1), (f-2) and (f-3). The case (f-1) was considered in Ref.\ \cite{kw04}, where the authors obtained the ratio between the decay rate of the true vacuum and that of the false vacuum. The radius and nucleation rate for all these cases were evaluated using the thin-wall approximation in Ref.\ \cite{ll03}.

We explain the $B$ with flat background. If $U_T \rightarrow 0$, then $\lambda_1 \rightarrow \lambda_2$ and the denominator $D^{1/2}$ becomes $(1+(\bar{\rho}_o/2\lambda_2)^2)$. As a result we obtain the same form of the radii for (f-4) and (f-6) nontrivially. It can be expected from Fig.\ \ref{fig:fig01}. The form is the same as that in Ref.\ \cite{bnu02}. It can be continuously matched at the case (f-5). The final form of $B$ is evaluated to be
\begin{equation}
B = B_{finite} + {\mathcal O}(U_T) \, ,
\end{equation}
where ${\mathcal O}(U_T)$ is vanishing and $B_{finite}$ is $\frac{-24\pi^2(1 + \frac{3\kappa S_o^2}{2\epsilon})}{\kappa^2\epsilon[1+(\frac{\bar{\rho}_o}{2\lambda_2})^2]^2} $ for (f-4), $\frac{-24\pi^2}{\kappa^3S_o^2}$ for (f-5), $\frac{-24\pi^2(1 + \frac{3\kappa S_o^2}{2\epsilon})}{\kappa^2\epsilon[1+(\frac{\bar{\rho}_o}{2\lambda_2})^2]^2}$ for (f-6), respectively. The form of the nucleation rate for the cases (f-4) and (f-6) turns out to be the same  nontrivially. For the case (f-4), we used the relation $(1-\frac{\kappa}{3}U_F\bar\rho^2)^{3/2} =(1-\frac{\kappa S_o \bar{\rho}}{2})^3$. For the case (f-6), we used the relation $(1-\frac{\kappa}{3}U_F\bar\rho^2)^{3/2} =(\frac{\kappa S_o \bar{\rho}}{2}-1)^3$. The nucleation rate can be also continuously matched among three cases. For all three cases, $B$ is negative. This is because the main contribution to $B$ comes from the $B_{out} < 0$, even though $B_{in}+B_{wall}>0$. In other words, $B=S^F - S^T < 0$, where $S^F <0$ and finite, and $S^T =0$.

We explain the nucleation rate for AdS background, which corresponds to the cases (f-7), (f-8), (f-9), (f-10), and (f-11).
As the size of the AdS background space goes to infinity which means $\eta_{max} \rightarrow \infty$, then $B_{out} \rightarrow + \infty$. The exponent of the nucleation rate is $B=B_{in} + B_{wall} + B_{out}$. $B_{out}$ gives the dominant contribution in $B$, i.e., $B \rightarrow + \infty$. In other words, $S^F <0$ and finite for the solutions, however $S^T \rightarrow -\infty $ for the AdS background which gives $B \rightarrow + \infty$. The probability for the nucleation is exponentially suppressed. In order to obtain the finite value of $B$, we need to employ the finite size $\tilde{\rho}^i_{max}$ cutoff for the initial space or the counter term to renormalize the action in AdS space.

\section{Summary and Discussion \label{sec4}}

We have studied the nucleation process of the false vacuum bubbles in Einstein gravity with a minimally coupled scalar field. The potential has one higher and the other lower vacuum state. The tunneling from the higher to the lower vacuum state, down-tunneling, is quite natural and always possible. The inverse tunneling from the lower to the higher, up-tunneling, does not always occur. The solutions exist only if gravity is taken into account. In this work, we investigated possible up-tunnelings and classified the types of false vacuum bubbles, which depends on the size of dS regions and the sign of the cosmological constant of the each vacuum region. We have shown the existence of new false vacuum bubbles by numerically solving the coupled equations for the gravity and the scalar field simultaneously.

The false vacuum solutions only with compact geometry are possible in Einstein gravity. The radius of regions and the nucleation probability are obtained as analytic computation using the thin-wall approximation. The solutions are as follows (See, Fig.\ \ref{fig:fig01}): (f-1) large dS bubble - small dS geometry, (f-2) half dS bubble - small dS geometry, (f-3) small dS bubble - small dS geometry, (f-4) large dS bubble - finite flat geometry, (f-5) half dS bubble - finite flat geometry, (f-6) small dS bubble - finite flat geometry, (f-7) large dS bubble - finite AdS geometry, (f-8) half dS bubble - finite AdS geometry, (f-9) small dS bubble - finite AdS geometry, (f-10) flat bubble - finite AdS geometry, (f-11) AdS bubble - finite AdS geometry. We interpret the nucleation process as tunneling events. The solutions could be expected by $\eta \rightarrow \eta_{max} -\eta$ from the case of a true vacuum bubble. The important thing is to show the existence of the false vacuum regions as the numerical solutions and obtain the nucleation rate. We expect that our solutions are nonsingular at $\eta_{max}$, because we are sure our numerical solutions being smooth at $\eta_{max}$. We demanded the conditions $d\Phi/d\eta \rightarrow 0$ and $\rho \rightarrow 0$ as $\eta \rightarrow \eta_{max}$ to ensure the regular solutions at $\eta = \eta_{max}$. If the initial background space has the finite size the tunneling interpretation will be meaningful.

In this work we also evaluated the nucleation rate of the false vacuum bubble solutions and the background. Our solutions with compact geometries have neither the surface term nor YGH boundary term, while the backgrounds for flat and AdS space have both two terms in Eq.\ (\ref{fe-action}). We obtained the finite probability using the thin-wall approximation for the cases (f-4), (f-5) and (f-6), i.e.\ the dS false vacuum bubbles with the flat background. It is not clear how to interpret the physical meaning of the negative value of the action in the present work.

For the case (f-7), (f-8), (f-9), (f-10) and (f-11), the diverging background subtraction causes the probability for the solutions being exponentially suppressed. We also correct the probability for the solution in the AdS background to be exponentially suppressed instead of be diverged in Ref.\ \cite{lllo10}. In order to obtain the finite value of $B$, we need to employ the finite size $\tilde{\rho}^i_{max}$ cutoff for the initial background space or the counter term to regularize the action in AdS space.

The negative sign is related to the fact that the Euclidean action for Einstein gravity is not bounded from below, which is known as the conformal factor problem in Euclidean quantum gravity \cite{ieqg, ieqg01}. Anyway, we can obtain the finite action for the solutions and the finite $B$ for the cases (f-4), (f-5), and (f-6). The important thing is the fact that, once the dS vacuum can be nucleated, the inflationary expansion could be eternal into the future and has the possibility of the self-reproduction.

For the nucleation of true vacuum bubbles, there exist the Great Divide representing the separation between true vacuum bubbles with a non-compact and those with a compact geometry of zero and negative cosmological constant. The solution with the compact geometry can be interpreted as a kind of domain wall production. For the false vacuum bubbles, there does not exist the Great Divide. We showed solutions only compact geometry are allowed.

We began our study on the false vacuum bubbles with the following three questions:
\begin{itemize}
\item Whether the universe could avoid a initial singularity or not?
\item Whether the universe came from something or nothing?
and
\item How could the universe be created with a very low entropy state?
\end{itemize}

We have had a question on avoiding the initial singularity problem and on the universe from nothing or something. We expect that if the inverse tunneling from the flat or AdS to dS was possible in the very early universe the whole universe could be complete in the past direction without the initial singularity and the universe could be created from something within the multiverse or the cosmic landscape scenario. If our solutions are applied to the scenarios, the concept on the terminal vacuum or sink could be modified. For the entropy issue, the scenario by the nucleation of a false vacuum bubble could also provide the beginning with a zero or very low entropy state. If the nucleation process could occur occasionally then the the entropy of the given background could be unbounded from above by the irreversable events.

We expect that the multiverse or the cosmic landscape scenario could resolve some of the puzzles including the cosmological constant problem. If the scenario has a huge number of different metastable and stable vacua, the study on the tunneling among different vacua could have physical significance and deserve further investigation. If the universe allows the successive tunneling process to lower vacuum or high vacuum state even it will take the extremely long time, there could be no Boltzmann brain problem \cite{also9, also91, also92}. The successive tunneling process could remedy the negative action occurred in a single tunneling event thanks to the adding of the contribution for the cases with the positive action. If our mechanism works in the whole universe, the universe starts with zero entropy and increases to the maximum value according to the inflationary scenario and black hole formations after dS regions creation. However the entropy of the universe is unbounded from above due to the uncleation in our framework. The measure problem including our results in the scenarios remains to be explored.

\section*{Acknowledgements}

We would like to thank Andrei Linde, Eduardo I.~Guendelman, Misao Sasaki, Hongsu Kim, Jiro Soda, Hyeong-Chan Kim, O-Kab Kwon, Inyong Cho, Wha-Keun Ahn, Dong-han Yeom, Seyen Kouwn, Daeho Ro, and Erick J.~Weinberg for helpful discussions and comments. We would like to thank Sang Pyo Kim, Mu-In Park, and Seoktae Koh for their hospitality during the Asia Pacific School on Gravitation and Cosmology (APCTP-NCTS-YITP Joint Program) in Jeju, Korea in 2013. We would like to thank Manu B.~Paranjape and Richard MacKenzie for their hospitality during our visit to Universit\'{e} de Montr\'{e}al. This work was supported by the National Research Foundation of Korea(NRF) grant funded by the Korea government(MSIP) (No. 2014002306). WL was supported by Basic Science Research Program through the National Research Foundation of Korea(NRF) funded by the Ministry of Education, Science and Technology(2012R1A1A2043908). We appreciate APCTP for its hospitality during completion of this work.

\end{document}